\documentclass[prb,twocolumn,showpacs]{revtex4}
\usepackage{graphicx,epsf}
\usepackage{bm, bbm}      

\newcommand{\rhobar}{\bar{\rho}}

\newcommand{\bq}{{\bf q}}
\newcommand{\bk}{{\bf k}}

\newcommand{\be}{{\bf e}}

\newcommand{\bR}{{\bf R}}

\newcommand{\eps}{\varepsilon}

\newcommand{\beq}{\begin{equation}}
\newcommand{\beqn}{\begin{eqnarray}}
\newcommand{\eeq}{\end{equation}}
\newcommand{\eeqn}{\end{eqnarray}}
\newcommand{\nn}{\nonumber}

\newcommand{\da}{\downarrow}
\newcommand{\ua}{\uparrow}

\newcommand{\SU}{{\rm SU}}

\newcommand{\Mmath}{\mathcal{M}}

\usepackage{color}

\begin{document}

\def\tende#1{\,\vtop{\ialign{##\crcr\rightarrowfill\crcr
\noalign{\kern-1pt\nointerlineskip}
\hskip3.pt${\scriptstyle #1}$\hskip3.pt\crcr}}\,}

\title{Theoretical Aspects of the Fractional Quantum Hall Effect in Graphene}
\author{M. O. Goerbig$^{1}$ and N. Regnault$^{2}$}

\affiliation{
$^1$Laboratoire de Physique des Solides, CNRS UMR 8502, Univ. Paris-Sud, F-91405 Orsay cedex, France\\
$^2$Laboratoire Pierre Aigrain, D\'epartement de Physique, ENS, CNRS, 24 Rue Lhomond, F-75005 Paris, France}

\begin{abstract}

We review the theoretical basis and understanding 
of electronic interactions in graphene Landau levels, in the limit 
of strong correlations. This limit occurs when inter-Landau-level excitations may be omitted because they belong
to a high-energy sector, whereas the low-energy excitations only involve the same level, such that the kinetic energy
(of the Landau level) is an unimportant constant. Two prominent effects emerge in this limit of strong electronic
correlations: generalised quantum Hall ferromagnetic states that profit from the approximate four-fold spin-valley
degeneracy of graphene's Landau levels and the fractional quantum Hall effect. Here, we discuss these effects in the
framework of an SU(4)-symmetric theory, in comparison with available experimental observations.

\end{abstract}
\maketitle

\section{Introduction}

The theory of non-interacting massless Dirac fermions in two spatial dimensions provides the framework which allows
for the understanding of most of graphene's low-energy electronic properties.\cite{antonioRev} 
At first sight, this may seem astonishing because the Coulomb interaction is strictly speaking of an intermediate
strength; indeed, within a typical Coulomb-gas argument, one compares the average interaction energy $e^2k_F/\epsilon$, 
at the characteristic
length $\sim k_F^{-1}$, in terms of the Fermi wave vector $k_F$ and the dielectric constant $\epsilon$ of the environment surrounding
the graphene sheet, to the kinetic energy $\hbar v_F k_F$ at the same length scale. The ratio between these energies yields 
the coupling constant
$\alpha_G=e^2/\hbar \epsilon v_F\simeq 2/\epsilon$, which is reminiscent of the fine-structure constant in 
quantum electrodynamics if one replaces the speed of light $c$ by the Fermi velocity $v_F$, i.e. the characteristic 
velocity of the electrons in a material. Because $v_F\simeq c/300$ in graphene, the \textit{graphene fine-structure constant} $\alpha_G$
is roughly 300 times larger than that ($\alpha=1/137$) of quantum electrodynamics. 

In view of the rather large coupling constant, one might expect to see correlation effects in graphene, which happen though to 
be sparse.\cite{kotov} Indeed, electronic instabilities are not only triggered by the (bare) coupling constant, but one also needs to take into 
account the density of states at the Fermi level in the discussion of such instabilities.\cite{mahan,GV} As a consequence of the 
linearity and the two-dimensional (2D) character of graphene electrons, the density of states, however, vanishes linearly with the
Fermi energy when approaching the limit of undoped (intrinsic) graphene, such that electronic instabilities are suppressed. In the
search of prominent correlation effects, one should therefore investigate situations in which the density of states in graphene is enhanced. A (logarithmically) diverging density of states is typically encountered at van Hove singularities due to saddle points in
the band dispersion. Van Hove singularities occur at extremely high energies ($\sim 3$ eV) in monolayer graphene and are
thus inaccessible with the help of field-effect doping. In contrast to monolayer graphene, they occur 
at rather small energies in AB-stacked bilayer graphene ($\sim 3$ meV),
\cite{MF06} which are only resolved at electronic densities below $10^{11}~\text{cm}^{-2}$. A promising system in this respect is
twisted bilayer graphene, where van Hove singularities at intermediate energies ($\sim 10...100$ meV) have been observed. \cite{vanHoveTwist} Such twists naturally occur in epitaxial graphene grown on the carbon face of the SiC crystal.\cite{deheer}

An alternative means of inducing a large density of states in graphene, and thus of increasing the role of electronic correlations, is 
to expose the sample to a strong perpendicular magnetic field $B$. In this case the electronic energy is quantised into highly 
degenerate Landau levels (LLs) at discrete energies $E_n$, around which the density of states $\rho(E)$ is strongly peaked,
$\rho(E)=gn_B \sum_n f(E-E_n)$, where $n_B=eB/h$ is the flux density measured in units of the flux quantum $h/e$ and $g$ takes
into account internal degrees of freedom, such as the four-fold spin-valley degeneracy ($g=4$) in graphene. The functions
$f(E-E_n)$, which are normalised to one, $\int dE\, f(E-E_n)=1$, become delta functions in the clean limit, which we assume in 
the theoretical discussion here. Each LL may thus be viewed, in this limit, as an infinitely flat energy band the density 
of states of which grows linearly with the magnetic field.

In this article, we review some effects due to the magnetic-field induced electronic correlations in graphene, in comparison with
the perhaps better-known 2D electron gas in semiconductor heterostructures. The probably most prominent one is the fractional 
quantum Hall effect (FQHE), which has recently been observed experimentally in the two-terminal\cite{du09,bolotin09} and
the four-terminal configuration.\cite{ghahari10,dean10} In contrast to the FQHE in GaAs heterostructures, the graphene FQHE
reflects a four-component structure\cite{toke07,GR07} that is inherited from the four-fold spin-valley degeneracy and that goes
along with particular magnetic properties described in the framework of SU(4) quantum Hall ferromagnetism.\cite{goerbigRev} 
The latter is also relevant 
in the discussion of interaction-induced integer quantum Hall effects (IQHE) at integer filling factors $\nu$ that do not belong to the 
``magic'' series $\nu=n_{el}/n_B=\pm 2,\pm 6, \pm 10, ...$, in terms of the carrier density $n_{el}$.

The article is organised as follows. In Sec. \ref{sec:Exp}, we review some of the experimental findings from the observations
of the IQHE in 2005 to the very recent ones of the FQHE in the four-terminal geometry, in 2010. After an introduction 
to the theoretical basics of graphene LLs in Sec. \ref{sec:Theo}, we discuss the SU(4)-spin-valley quantum Hall ferromagnetism
in Sec. \ref{sec:QHFM} and the SU(4) theory of the FQHE in Sec. \ref{sec:FQHE}.

\section{Experimental Situation}
\label{sec:Exp}

\subsection{Relativistic integer quantum Hall effect}

A milestone experiment in graphene research was the observation  in 2005 of a particular -- relativistic -- IQHE in graphene, when 
changing either the electronic density via the electric-field effect at a fixed magnetic field or when varying the field
at fixed electronic 
density.\cite{novoselov05,zhang05} 
The samples used in these magnetotransport measurements were obtained with the help of the exfoliation technique,\cite{exfol}
and the effect has later (in 2009) been confirmed in epitaxial graphene samples,\cite{epitaxIQHE} which have also been proven to be 
promising for metrological means because of a high-precision (with an error bar on the order of $10^9$) 
Hall-resistance quantisation.\cite{tza}
Whereas the effect has the same signature -- a plateau in the Hall resistance accompanied by a 
vanishing longitudinal resistance -- as that in conventional 2D electron systems, it occurs at unusual filling factors, 
\beq\label{eq:RQHE}
\nu=\pm 2(2n+1)= \pm 2, \pm 6, \pm 10, ..., 
\eeq
and reflects the relativistic nature of the charge carriers in graphene. Indeed, the
two possible signs $\pm$ reflect the presence of a conduction band ($+$ for ``particles'') that touches the valence band ($-$ for
``anti-particles'' on the hole-doped side). The filling-factor steps in units of four between successive plateaus may easily be 
understood as a consequence of the four-fold spin-valley degeneracy, which was not resolved in these first experiments and that
yields four copies of each LL. The offset of $\pm 2$ in the plateau series
(\ref{eq:RQHE}) is a consequence of \textit{relativistic LL quantisation} that yields a LL spectrum
\beq\label{eq:LL}
E_{\lambda,n}=\hbar\frac{v_F}{l_B}\sqrt{2n},
\eeq
where $l_B=\sqrt{\hbar/eB}\simeq 26/\sqrt{B\text{[T]}}$ nm is the magnetic length, and the integers $n$ label the levels
in the conduction band ($\lambda=+$) or in the valence band ($\lambda=-$). The most
prominent feature of the level spectrum (\ref{eq:LL}), apart from its square-root dispersion with the magnetic field and with $n$, is 
the presence of a zero-energy LL for $n=0$. It is this level that is responsible for the offset $\pm 2$ in the series (\ref{eq:RQHE})
because it is only half-filled at zero doping, $\nu=0$. This means that the condition for the IQHE, namely a set of completely 
filled LLs with a topmost filled level separated by a gap from the lowest unoccupied LL, is not fulfilled at $\nu=0$, but only at 
$\nu=2$ (for electron doping) or $\nu=-2$ (for hole doping), as a consequence of the four-fold spin-valley degeneracy of the $n=0$ LL.

\subsection{Additional plateaus at integer fillings}
\label{sec:FMexp}

In 2006, one year after the discovery of the graphene IQHE, novel high-field plateaus have been observed at $\nu=0,\pm 1$ and $\pm 4$ 
that do not belong to the series (\ref{eq:RQHE}).\cite{zhang06} These additional states indicate that the spin-valley degeneracy 
in the $n=0$ LL is completely lifted, whereas in $n=1$ it is only partially lifted -- if it were fully lifted in the latter case, 
one would also expect an IQHE at $\nu=\pm 3$ and $\pm 5$. The explanations which have been given for the spin-valley 
degeneracy lifting fall into two classes: (1) extrinsic or (2) intrinsic, i.e. interaction-induced, effects. The simplest 
extrinsic effect is certainly the Zeeman effect that would lift the spin degeneracy, such that each four-fold degenerate LL is split into
two (valley-degenerate) spin branches separated by an energy scale of $\Delta_Z\simeq 1.2 B\text{[T]}$ K. A more subtle extrinsic 
effect, as a consequence of electron-phonon coupling, is capable of lifting the valley degeneracy in the zero-energy LL in form of
the generation of a mass gap in the level spectrum.\cite{FL,nomura09,mudry10} The coupling to an out-of-plane phonon can yield 
a Peierls-type distortion and can thus break the inversion symmetry of the lattice.\cite{FL} More recently an inplane Kekul\'e distortion
has been investigated that couples the two different valleys.\cite{nomura09,mudry10} In contrast to the out-of-plane distortion, the
latter mechanism yields a mass term (and thus a valley splitting) that does not depend on the coupling to the substrate and that 
has been evaluated to be roughly $\Delta_{kek}\simeq 2 B\text{[T]}$ K. Notice that the linear $B$-field dependence simply reflects 
the fact that the coupling is proportional to the density of states $\rho(E)\propto B$, which scales linearly with the magnetic field, as mentioned in the introduction.

The second class of degeneracy-lifting effects contains mechanisms that are triggered by the Coulomb interaction between the electrons
and that are discussed in more detail in Sec. \ref{sec:QHFM}. One effect is the so-called \textit{magnetic catalysis}, which has been investigated before the discovery of the IQHE in graphene, in the context of Dirac fermions.\cite{khvesh01}
The mechanism consists of a mass-gap generation that yields the
same level spectrum (and thus a valley-degeneracy lifting) as that due to the above-mentioned Peierls-type distortions, but it
is dynamically generated by the electron-electron interactions themselves.\cite{khvesh01,Mcata} The mass term plays the role
of an order parameter that has been identified with exciton condensation. Independently, quantum-Hall ferromagnetism, both in the spin 
and in the valley channel, has been proposed in 2006 as a possible route to understanding the additional 
plateaus.\cite{nomura06,GMD,AF06,YDSMD06} Quantum-Hall ferromagnetism is an exchange effect, where the electron-electron interaction
is minimised by the formation of a maximally antisymmetric orbital wave function, accompanied by a maximally symmetric valley-spin
part. The effect is particularly efficient in LLs because the latter may be viewed as infinitely flat bands -- the polarisation 
of the spin and the valley \textit{pseudospin} is therefore not accompanied by a cost in kinetic energy. Whereas quantum-Hall 
ferromagnetism is capable of generating a transport gap, and thus an IQHE, at all integer filling factors that do not belong to 
the series (\ref{eq:RQHE}),\cite{arovas} 
a mass gap resulting from a lattice distortion or magnetic catalysis can only lift the valley degeneracy
in the zero-energy LL $n=0$. However in both cases, magnetic catalysis and quantum-Hall ferromagnetism, the gap scales with the
typical interaction energy $e^2/\epsilon l_B\propto \sqrt{B}$. 

\begin{center}
\begin{table}[htbp]
{
\begin{tabular}{|c||c|c|}
\hline
energy & value for arbitrary $B$ & for $B=25$ T\\
\hline \hline
$\Delta_{Z}$ & $1.2 B{\rm [T]}$ K &  $30$ K \\ \hline
$\Delta_{kek}$ & $2 B{\rm [T]}$ K &  $50$ K \\ \hline\hline
$e^2/\eps l_B$ (vacuum) & $139\sqrt{B{\rm [T]}}$ K &  $694$ K
\\ \hline
$e^2/\eps l_B$ (on SiO$_2$) & $104\sqrt{B{\rm [T]}}$ K &  $521$ K
\\ \hline
$e^2/\eps l_B$ (on h-BN) & $109\sqrt{B{\rm [T]}}$ K &  $543$ K
\\ \hline
$e^2/\eps l_B$ (on SiC) & $71\sqrt{B{\rm [T]}}$ K &  $355$ K\\
\hline
\end{tabular}}
\label{Tab:Zeeman}
\caption{\footnotesize Tab. I. Energy scales for spin-valley degeneracy lifting in graphene LLs. The first two lines show the energy scales associated with extrinsic effects (Zeeman effect and Kekul\'e-type lattice distortion, 
$\Delta_Z$ and $\Delta_{kek}$, respectively), which are proportional to $B$. 
Below are shown the interaction-energy scales ($\propto \sqrt{B}$), different substrates taking into account 
both the dielectric constant of the substrate and RPA contributions from inter-band processes. 
}
\end{table}
\end{center}

The discussed energy scales are summarised in the table above. The interaction energy scales depend on the dielectric constant 
$\epsilon$ of the environment, which consists in a typical experimental situation on the substrate [SiO$_2$, hexa boron nitride
(h-BN) or SiC for epitaxial graphene] on one side and air (vacuum) on the other one. The third line (vacuum) indicates the energy
scale for freestanding graphene. A recent theoretical study proposes to engineer the short-range part of interaction potential
via a partial screening with a dielectric medium at a finite distance from the graphene sheet.\cite{Princeton}
Notice that in all cases, we have taken into account screening due to the completely filled valence 
band, which yields $\eps=\epsilon(1+\pi\alpha_G/2)$ within the random-phase approximation.\cite{GGV} One notices from these energy
scales that in all cases the Coulomb interaction sets the leading energy scale and should thus be considered first in the discussion
of the spin-valley degeneracy lifting, whereas extrinsic effects are subordinate. As it is discussed below in Sec. \ref{sec:QHFM}, 
the extrinsic effects are cooperative with quantum-Hall ferromagnetism in the sense that they orient the interaction-induced 
spin-valley magnetisation in a particular direction.

\subsection{Fractional quantum Hall effect in graphene}

In the previous subsection, we have argued that the appearance of IQHE plateaus, which do not match the series (\ref{eq:RQHE}), could in 
principle be understood without invoking electron-electron interactions -- extrinsic effects could
be responsible for these plateaus, although this is unlikely in
view of the different energy scales involved. Clear evidence for interaction-induced phases in graphene LLs has been found in 2009 with
the first observations of the FQHE at $\nu=1/3$
in suspended graphene.\cite{du09,bolotin09} One notices that it took roughly twice as long in graphene
between the observation of the IQHE and the FQHE as compared to conventional 2D electron systems, where the IQHE was observed in 
1980,\cite{KvK} whereas the FQHE was discovered in 1982.\cite{TSG} The necessary mobility increase of graphene samples, which
is required for the observation of the FQHE, could already be achieved in 2008 in current-annealed suspended samples,
where mobilities in the $100\, 000~\text{cm}^2/\text{Vs}$ range have been reported.\cite{du08} However, 
it turned out to be an experimental challenge to obtain samples with working and sufficiently separated electronic contacts. 

The  above-mentioned transport measurements, which revealed the FQHE, were indeed performed in the two-terminal configuration, where the
same contacts used as source and drain serve for the resistance measurement. It is therefore not possible to perform simultaneously
a measurement of the longitudinal and the Hall resistance, but both are superposed, and sophisticated conformal mappings are necessary
to separate the two components.\cite{ConfMap} However, because of the vanishing longitudinal component in the case of the FQHE,
the two-terminal resistance is then determined by the quantised ``Hall'' resistance and therefore reveals the characteristic plateau.

These first observations have since been confirmed in the more robust four-terminal configuration, which allows for a simultaneous 
measurement of and thus a clear distinction between the longitudinal and the Hall resistances. Two experiments were reported in 2010, one
in a suspended graphene sample\cite{ghahari10} and another one in graphene on an h-BN substrate\cite{dean10} 
that allows for a mobility increase upon current annealing that is in the same range as (though somewhat lower than)
that in suspended graphene.\cite{dean10a} In Ref. \onlinecite{ghahari10} the activation gap could be determined and agrees 
rather well with that $0.05 ... 0.1 e^2/\eps l_B\sim 7...14 \sqrt{B{\rm [T]}}$ K one expects\cite{AC06,toke06} for the 
polarised Laughlin state at $\nu=1/3$.\cite{laughlin} In graphene on a h-BN substrate, most members of the $1/3$ FQHE family in 
$n=0$ (at $\nu=\pm 1/3,\pm 2/3$, and $\pm 4/3$) and all in $n=1$ (at $\nu=\pm 7/3,\pm 8/3, \pm 10/3$, and $\pm 11/3$) could be resolved
to great accuracy.\cite{dean10} 
Whereas the absence or relative weakness of the $\pm 5/3$ member of the $1/3$-family in the zero-energy LL
remains to be understood, it clearly corroborates the approximate SU(4) symmetry due to the four-fold spin-valley degeneracy underlying
the FQHE in graphene,\cite{toke07,GR07} as discussed in more detail from the theoretical point of view in Sec. \ref{sec:FQHE}.
Another relevant finding in graphene on an h-BN substrate is that of additional IQHE plateaus at $\nu=\pm 3$ and $\pm 5$,\cite{dean10}
which indicate a full spin-valley degeneracy lifting not only in $n=0$ but also in $n=1$. Whereas these additional plateaus cannot 
be explained in the framework of a mass-gap generation, either by a lattice distortion or interaction-induced magnetic catalysis, 
it is expected from the formation of quantum-Hall ferromagnetic states, as mentioned above.

\section{Theoretical Understanding}
\label{sec:Theo}

From the theoretical point of view, both the FQHE and quantum-Hall ferromagnetism require three essential ingredients: (1) 
infinitely flat and highly-degenerate energy bands (here in the form of LLs the degeneracy of which is characterised 
by the flux density $n_B$); (2) the Aharonov-Bohm effect, which yields a geometric phase to paths in the 2D plane and that
induces the so-called magnetic translation group; and (3) sufficiently short-range interactions.\footnote{Contrarily to the 
ususal case of electrons without a magnetic field, the Coulomb interaction is considered as a short-range interaction.}

In order to make transparent the above statements, we consider a single LL that is sufficiently well separated in energy from
its adjacent levels. This condition is fulfilled when the LL spacing $\Delta_n=\sqrt{2}\hbar (v_F/l_B)(\sqrt{n+1}-\sqrt{n})\simeq
\hbar v_F/\sqrt{2n}l_B\simeq 200\sqrt{B\text{[T]}/n}$ K is larger than the impurity broadening of the levels. Furthermore,
because this energy scale is much larger than the extrinsic spin-valley symmetry-breaking effects (see Tab. I), we may consider
each LL as four-fold degenerate, in addition to the orbital degeneracy given by the flux density $n_B$. The electron-electron
interactions may then be separated into a low-energy and a high-energy part. The latter consists of interaction-induced inter-LL
transitions at the characteristic energy scale $\Delta_n$, whereas the low-energy part consists of intra-LL excitations, in which
case the kinetic energy (set by the scale $\hbar v_F/l_B$) effectively drops out of the problem. The 
low-energy interaction Hamiltonian (in reciprocal space) may then be written as
\beq\label{eq:intHam}
H_n=\frac{1}{2}\sum_{\bq}v(q)\rho_n(-\bq)\rho_n(\bq),
\eeq
where $v(q)=2\pi e^2/\eps q$ is the Fourier-transformed Coulomb interaction potential, and the Fourier components 
of the density operator $\rho_n(\bq)$ take into account only states within the $n$-th LL. Apart from a form factor 
$\mathcal{F}_n(q)$ that
takes into account the overlap between electronic wave functions in the $n$-th LL and that may be absorbed into an effective
interaction potential,\cite{goerbigRev} the (projected) density operator 
\beq
\rhobar(\bq)\equiv \frac{\rho_n(\bq)}{\mathcal{F}_n(\bq)}=\sum_{j=1}^{N} e^{-i\bq\cdot\bR_j},
\eeq
is a sum of the one-particle density operators 
$\rhobar_j(\bq)=\exp(-i\bq\cdot\bR_j)$
for each of the $N$ electrons in the $n$-th LL. Here, $\bR_j=(X_j,Y_j)$ is the operator that describes the centre of 
the cyclotron motion (called \textit{guiding centre})
of the $j$-th electron. Because it is a constant of motion, it does not connect states in different 
LLs, in agreement with the construction of the model (\ref{eq:intHam}). Furthermore, the components of $\bR_j$ do not
commute, 
\beq\label{eq:comm}
[X_j,Y_{j'}]=il_B^2\delta_{j,j'},
\eeq
which is a manifestation of the above-mentioned Aharonov-Bohm effect. Indeed, one sees from the commutation relations that
$X_j$ and $Y_j$ are conjugate variables, such that $X_j$ generates a translation in the $-y$-direction, whereas 
$Y_j$ generates one in the $x$-direction. As a consequence the electron, when moving on a closed path around an area $\Sigma$,
picks up an Aharonov-Bohm phase $\varphi=\Sigma/l_B^2=2\pi \phi/\phi_0$, where $\phi=B\Sigma$ is the flux in the area $\Sigma$.
The commutation relations (\ref{eq:comm}) furthermore induce the commutation relations 
\beq\label{eq:commP}
[\rhobar(\bq),\rhobar(\bk)]=2i\sin\left(\frac{q_xk_y-q_yk_x}{2}l_B^2\right)\rhobar(\bq+\bk)
\eeq
for the projected density operators, such that their Heisenberg equations of motion $i\hbar\dot{\rhobar}(\bq)=[\rhobar(\bq),H]$
become highly non-linear.

Another consequence of the commutation relations (\ref{eq:comm}), which allow for the introduction of 
harmonic-oscillator ladder operators
$b_j=(X_j + iY_j)/\sqrt{2}l_B$ and $b_j^{\dagger}=(X_j - iY_j)/\sqrt{2}l_B$, with $[b_j,b_{j'}^{\dagger}]=\delta_{j,j'}$,
is the representation of $n=0$-LL states in terms of analytic functions
\beq
\phi_m(z_j,z_j^*)\sim \left(b_j^{\dagger}\right)^m e^{-|z_j|^2/4}\sim z_j^m  e^{-|z_j|^2/4},
\eeq
where we have defined the complex position $z_j=(x_j-iy_j)/l_B$ of the $j$-th particle in the 2D plane and where we 
have omitted the normalisation constant in the expressions. Whereas this statement
is, strictly speaking, true only in the $n=0$ LL, one may nevertheless use a representation of states in other LLs in terms
of analytic functions if one interprets $v_n(q)=v(q)|\mathcal{F}_n(q)|^2$ as an effective interaction potential that mimics the
$n$-th LL while considering the projected density $\rhobar(\bq)$ as one of $n=0$. This assumption is justified by the fact that
the commutation relations (\ref{eq:commP}) do not explicitly depend on the LL index. 

To summarise this theoretical introduction of the model, one notices the following important points. 
\begin{itemize}

\item The above arguments are valid for any type of Landau quantisation and not restricted to the relativistic one in graphene. From
this point of view, graphene and its FQHE are not so different from the FQHE in semiconductor heterostrucures. 

\item The specificity of graphene and relativistic LL quantisation is revealed rather in the form factors $\mathcal{F}_n(q)$,
which take into account the wave function overlaps in a particular LL $n$ and that happen to be different from that in non-relativistic
2D electron systems, as a consequence of the spinorial structure of the electronic wave functions in graphene.

\item Another specificity of graphene is its four-fold spin-valley degeneracy. The Coulomb interaction naturally commutes with the
electronic spin, whereas this is \textit{a priori} not the case 
for the valley pseudospin -- indeed, the two different valleys can be coupled by
scattering due to short-range components of the Coulomb potential. However, one may show that this inter-valley coupling is suppressed
by a factor of $a/l_B\sim 0.005 \sqrt{B\text{[T]}}$ because of the reciprocal-space distance $\sim a^{-1}$ between the two 
valleys,\cite{GMD,AF06} such that the Coulomb interaction in graphene LLs may be viewed as approximately SU(4)-symmetric.

\item A general $N$-particle wave function in graphene LLs must therefore be described by an analytic polynomial in all
particle coordinates $z_k^{(j)}$, where the superscript $(j)$ indicates one of the four-spin valley components $(K,\ua)$,
$(K,\da)$, $(K',\ua)$, or $(K',\da)$.

\end{itemize}

\subsection{Interaction-induced integer quantum Hall effect}
\label{sec:QHFM}

\begin{figure}
\centering
\includegraphics[width=5.5cm,angle=0]{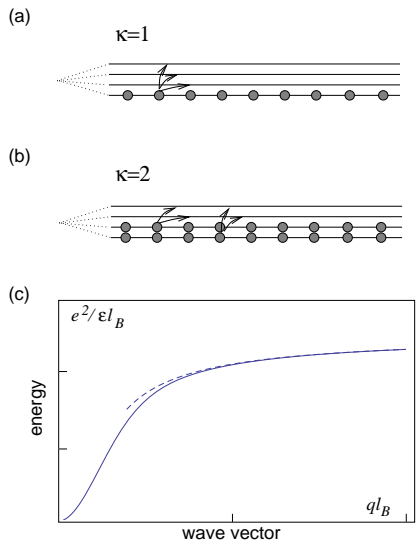}
\caption{\footnotesize{Spin-valley waves (Goldstone modes) in graphene LLs. 
\textit{(a)} For one filled spin-valley subbranch ($\kappa=1$), there are 
three Goldstone modes connecting the filled subbranch with the empty ones.
\textit{(b)} For two filled subbranches ($\kappa=2$), 
four Goldstone modes connect the completely filled subbranches with the empty ones. 
\textit{(c)} Dispersion relation of the spin-valley waves (continuous line). At $ql_B\ll 1$, the energy disperses as 
$E_{\bq}\propto q^2$, 
as one expects for magnons. At $ql_B\gg 1$, the dispersion relation saturates and may be approximated by the energy to create 
a well-separated electron-hole pair, with a Coulomb attraction between the electron and the hole, 
$E_{\bq}\sim [2\sqrt{\pi/8}-1/q l_B]e^2/\epsilon l_B$ (dashed line).
}}
\label{fig:SW}
\end{figure}

A first manifestation, the understanding of which 
turns out to be instructive also for the FQHE, of Coulomb interactions in graphene LLs is the formation
of SU(4) quantum-Hall ferromagnetic states at integer filling factors that do not correspond to completely filled LLs described by 
the series (\ref{eq:RQHE}). As already mentioned above, a repulsive interaction such as the Coulomb potential favours orbital 
wave functions with nodes when two particles approach each other. These nodes are naturally built in in the usual Slater determinants
for $\kappa$ completely filled spin-valley branches of the last (partially) occupied LL,
\beq\label{eq:QHFM}
|{\rm FM}\rangle = \prod_{j=1}^{\kappa}\prod_{m=0}^{N_B-1} c_{m,j}^{\dagger}|{\rm vac}\rangle,
\eeq
where $c_{m,i}^{\dagger}$ creates an electron in the component $j$ in the state corresponding to the LL wave function $z^m$ and
$|{\rm vac}\rangle$ denotes the fermion vacuum. For illustration, we consider the state (\ref{eq:QHFM}) in the zero-energy LL $n=0$
and omit the LL index at the fermion operators, keeping in mind that the generalisation to other LLs is straight-forward, as discussed
above. 

Naturally, the state (\ref{eq:QHFM}) is the ground state of a model with no interactions
if the filled states have a lower one-particle energy than the empty ones, for example in the presence of a Zeeman effect. Here, however,
we consider these effects to be 
absent, and we do therefore not specify whether the occupied branches are particular spin or valley states.
The state (\ref{eq:QHFM}) therefore breaks the SU(4) spin-valley symmetry, which is respected by the interaction model. As a consequence
of this symmetry breaking, the quantum-Hall ferromagnet (\ref{eq:QHFM}) has low-energy excitations in form of Goldstone modes the 
energy of which vanishes in the small-wave-vector limit. These Goldstone modes are spin-valley waves that connect the different possible 
ground states, which are obtained by relabeling the occupied subbranches. They are depicted in Fig. \ref{fig:SW} for $\kappa=1$ (a) and
$\kappa=2$ (b) completely filled subbranches -- the case of $\kappa=3$ 
filled subbranches is particle-hole-symmetric to $\kappa=1$. For $\kappa=1$, the
Goldstone modes are three-fold degenerate [see Fig. \ref{fig:SW}(a)] whereas for $\kappa=2$ there are four different types 
[Fig. \ref{fig:SW}(b)]. 

\subsubsection{Spin-valley waves}

The different spin-valley waves may be obtained by application of the operator
\beq
\rhobar_{ij}(\bq)=\sum_{m,m'}\left\langle m\left|e^{-i\bq\cdot\bR}\right|m'\right\rangle c_{m,i}^{\dagger}c_{m',j}
\eeq
on the state $|{\rm FM}\rangle$,
where $\bR$ is the one-particle operator associated with the guiding centre, as discussed above, and $j$ denotes 
an occupied spin-valley component, whereas $i$ corresponds to an unoccupied one. The state $\rhobar_{ij}(\bq)|{\rm FM}\rangle$
may also be viewed as a superposition of particle-hole excitations, where $\bq$ is the wave vector of the excitation. As a 
consequence of the magnetic translation algebra, generated by the commutation relations (\ref{eq:comm}), this wave vector is 
proportional to the distance $\Delta \bR=\bR-\bR'$ between the guiding centre $\bR$ of the electron and that $\bR'$ of the hole,
\beq\label{eq:wvgc}
\bq=\Delta \bR\times \be_z/l_B^2.
\eeq
The energy spectrum of the spin-valley waves may be obtained by evaluating the Hamiltonian (\ref{eq:intHam}) in the state 
$\rhobar_{ij}(\bq)|{\rm FM}\rangle$, and one obtains
\beqn\label{eq:SWdisp}
\nn
E_{\bq} &=& \langle {\rm FM}| \rhobar_{ij}(-\bq) H_n \rhobar_{ij}(\bq) - H_n|{\rm FM}\rangle\\
&=& 2
\sum_{\bk}v_n(\bk)\sin^2\left(\frac{q_xk_y-k_xq_y}{2}l_B^2\right),
\eeqn
or explicitly, in the zero-energy LL $n=0$,\cite{AF06,YDSMD06,KH,doretto}
\beq\label{eq:SWn0}
E_{\bq} = \sqrt{\frac{\pi}{2}}\frac{e^2}{\epsilon l_B}\left[1-e^{-q^2l_B^2/4}I_0\left(\frac{q^2l_B^2}{4}\right)\right],
\eeq
which is plotted in Fig. \ref{fig:SW}(c). In the last expression, which is independent of the number $\kappa$ of filled spin-valley 
branches and which thus indicates that the spin-valley waves are degenerate for all values of $\bq$,
$I_0(x)$ is a modified Bessel function. The limits of the dispersion (\ref{eq:SWn0}) are transparent; for small values of
the wave vector $ql_B\ll 1$, one obtains the usual $q^2$ dispersion expected for spin-wave-type modes,
\beq\label{eq:SWsmallQ}
E_{q\rightarrow 0} = \frac{\rho_s}{2} q^2l_B^2,
\eeq
in terms of the spin stiffness
\beq\label{eq:spinstiff}
\rho_s=\frac{1}{4\pi}\sum_{\bk}v_0(\bk)|\bk|^2l_B^2 =\frac{1}{16\sqrt{2\pi}}\frac{e^2}{\epsilon l_B}.
\eeq
In the opposite limit, the dispersion may be understood in terms of the energy of a spatially well-separated electron-hole pair. 
The energy to add an electron (or a hole) to the state (\ref{eq:QHFM}) is just given by the exchange energy, 
$E_x=\sqrt{\pi/8}(e^2/\epsilon l_B)$. The value at which the dispersion (\ref{eq:SWn0}) saturates is indeed twice the exchange
energy. Furthermore, the electron and the hole with opposite charge interact via the Coulomb attraction
\beq
\frac{e^2}{\epsilon |\Delta\bR|}=\frac{e^2}{\epsilon q l_B^2},
\eeq
as a consequence of the connection (\ref{eq:wvgc}) between the wave vector and the distance between the guiding centre of the electron
and that of the hole. As depicted by the dashed line in Fig. \ref{fig:SW}(c), the spin-valley-wave dispersion is well approximated
by the sum of these two terms,
\beq
E_{ql_B\gg 1}\simeq \left[2\sqrt{\frac{\pi}{8}} - \frac{1}{ql_B}\right]\frac{e^2}{\epsilon l_B}.
\eeq

To summarise the picture of SU(4) quantum Hall ferromagnetism and the associated spin-valley-wave modes, we first mention
that the polarised state (\ref{eq:QHFM}) may be obtained simply as a consequence of the Coulomb repulsion between the electrons
without the need of explicit spin-valley symmetry-breaking terms, such as the Zeeman effect. The state is stable because the dispersion
of the collective excitations is gapped for any non-zero value of the wave vector, and the addition of an electron (or a hole) is
associated with an energy cost given by the exchange energy, which is much larger than the external symmetry-breaking fields (see
Table 1). The role of such external terms is then reduced to a simple orientation of the interaction-induced spin-valley
magnetisation, similarly to a usual (spin) ferromagnet placed into a magnetic field that orients its magnetisation in
the direction of the field. 

In the following paragraph, we argue that there are lower-energy elementary excitations, in form of skyrmions, than such
additional electron with a flipped spin or valley-pseudospin. However, their energy is also determined by the interaction-energy
scale $e^2/\epsilon l_B$, such that the overall picture remains unaltered.

\subsubsection{SU(4) skyrmions}

In the previous paragraph, we have considered the elementary excitation to be a simple additional electron (or hole) that
is added into an unoccupied spin-valley component in the quantum-Hall ferromagnet (\ref{eq:QHFM}). Its energy is then simply
given by the exchange energy $E_x=\sqrt{\pi/8}(e^2/\epsilon l_B)$. However, it turns out to be energetically favourable for this
additional particle to be dressed by a local deformation of the SU(4)-ferromagnetic background, such as to lower the energy
cost due to the opposite spin orientation of the particle with respect to the background. This dressed particle is called 
\textit{skyrmion} and carries a topological charge in addition to its electric one. For a simple SU(2) spin ferromagnet,
this topological charge may be
viewed as the number of times the (normalised) local magnetisation wraps, when exploring the 2D plane,
the Bloch sphere the points of which represent the orientation of the magnetisation. The SU(4) spin-valley case is more complicated
and requires the introduction of two additional Bloch spheres (one for the valley-pseudospin and one for the two angles
that describe the entanglement between the spin and the valley-pseudospin),\cite{DGLM} but the picture is essentially the same. 

The topological charge $Q_{top}$, which is a positive or negative integer, determines the energy of the skyrmion 
excitation,\cite{sondhi,moon}
\beq\label{eq:SkyrmEn}
E_{sk}=4\pi \rho_s |Q_{top}|=\frac{1}{2}\sqrt{\frac{\pi}{8}}\frac{e^2}{\epsilon l_B}|Q_{top}|
\eeq
in terms of the spin stiffness (\ref{eq:spinstiff}). One thus notices that the energy to create a skyrmion with charge $Q_{top}=\pm 1$
is half of that to create a simple (undressed) electron in an unoccupied spin-valley component. Dressing this additional electron
by a topological spin-valley texture therefore lowers the energy in $n=0$ by a factor of 2. The energy gain is less in the LLs $n\neq 0$,
but it remains positive for $n=1$ and $n=2$, whereas in even higher LLs it is no longer energetically favourable to dress the 
additional charge by creating skyrmions.\cite{toke06,YDSMD06}

We finally mention that the skyrmion is generically larger in size than an undressed electronic excitation ($\sim l_B$). The size 
of the skyrmion is indeed determined by a competition between the Coulomb (exchange) interaction, which favours large skyrmions
to maintain locally the ferromagnetic order, and external symmetry-breaking terms that, even if they are small, have a
tendency to lower the number of reversed spins or valley-pseudospins and thus to lower the skyrmion size. Indeed, the skyrmion 
radius scales as\cite{sondhi,moon}
\beq
\xi \sim \sqrt{\frac{e^2/\epsilon l_B}{\Delta}}l_B,
\eeq
where $\Delta$ represents a generic spin-valley symmetry-breaking terms, such as the Zeeman effect or that arising from a spontaneous
lattice distortion discussed in Sec. \ref{sec:FMexp}.

\subsection{SU(4) fractional quantum Hall effect}
\label{sec:FQHE}

In the previous section, we have argued that electron-electron interactions are responsible for the formation of additional plateaus 
at integer filling factors that do not correspond to the series (\ref{eq:RQHE}), as a consequence of the formation of maximally
polarised quantum-Hall ferromagnetic states. These considerations turn out to be helpful also in the understanding of the 
four-component FQHE. If we consider, e.g., the SU(4) ferromagnetic state at $\nu=\pm 1$, its orbital wave function may be written
in terms of the completely anti-symmetric Slater determinant
\beq\label{eq:WFnu1}
\phi \left(\left\{z_k\right\}\right) = \prod_{k<l}^N\left(z_k-z_l\right) e^{ - \sum_{k=1}^N|z_k|^2/2},
\eeq
in terms of the complex coordinates $z_k$ of the electron in units of $l_B$, regardless of the spin-valley component they belong to. As 
a consequence of the anti-symmetry of this orbital wave function, the associated spin-valley wave function must be completely
symmetric, i.e. precisely ferromagnetic, such as to fulfil the anti-symmetry requirement for fermionic $N$-particle wave functions. 
Wave function (\ref{eq:WFnu1}) is the simplest example of Laughlin's wave function\cite{laughlin}
\beq\label{eq:Laughlin}
\phi_m^{L}\left(\left\{z_k\right\}\right) = \prod_{k<l}^N\left(z_k-z_l\right)^m e^{ - \sum_{k=1}^N|z_k|^2/2},
\eeq
which describes FQHE states at filling factors $\nu=1/m$. Indeed, a power counting of the terms in the polynomial indicates that
the largest power of an arbitrarily chosen particle component $z_k$ is $M=m(N-1)$. As we have already mentioned in the first
part of this section, this power is delimited by the number of flux quanta threading the 2D system, such that $M=N_B-1$,
and one obtains, in the thermodynamic limit, the relation 
\beq\label{eq:fillM}
m=\frac{N_B}{N}=\frac{n_B}{n_{el}}=\frac{1}{\nu},
\eeq
i.e. the exponent $m$ in Laughlin's wave function determines the filling factor. For odd values of $m$ -- remember 
from the previous discussion that $m$ must be an integer to match the analyticity condition for wave functions in the LL $n=0$ --
the same symmetry arguments apply as for the wave function (\ref{eq:WFnu1}). It is a fully anti-symmetric orbital wave function,
and the spin-valley part must therefore be completely symmetric, such that Laughlin's wave function (\ref{eq:Laughlin}) 
represents a fully polarised spin-valley ferromagnet. 

Similarly to Laughlin's wave function, the theory of composite fermions (CF)\cite{jain89} can be extended to include
the SU(4) internal degree of freedom.\cite{toke07} Still, the main
physical consequences of this additional symmetry can be captured within
the simpler framework of Halperin wave functions.\cite{halperin} Such states have been
introduced in 1983, soon after Laughlin's original work, to take into account the electronic spin and to
describe non-fully polarized FQHE states. This set 
of wave functions is readily generalised to the four-component case in graphene\cite{GR07}
\beq\label{eq:GenHalp}
\psi_{m_1,...,m_4;n_{ij}}^{\SU(4)}
=\phi_{m_1,...,m_4}^L \phi_{n_{ij}}^{inter},
\eeq
in terms of the product 
\beq
\phi_{m_1,...,m_4}^L=\prod_{j=1}^4 \prod_{k_j<l_j}^{N_j}\left(z_{k_j}^{(j)}-z_{l_j}^{(j)}\right)^{m_j}e^{-\sum_{j=1}^{4}
\sum_{k_j=1}^{N_j}|z_{k_j}^{(j)}|^2/4}
\eeq
of Laughlin wave functions for the four spin-valley components
and the term
\beq
\phi_{n_{ij}}^{inter}=\prod_{i<j}^{4}\prod_{k_i}^{N_i}\prod_{k_j}^{N_j}
\left(z_{k_i}^{(i)}-z_{k_j}^{(j)}\right)^{n_{ij}},
\eeq
which describes inter-component correlations. Here, $z_{k_j}^{(j)}$ is the complex coordinate of a particle in the component 
$j$, and $N_j$ is the total number of $j$-type particles. As in the case of Laughlin's wave function, the power-counting argument
relates the exponents $m_j$ and $n_{ij}$ to the \textit{component filling factors} $\nu_j=N_j/N_B$. 
Indeed, for an arbitrarily chosen component $j$, the maximal exponent is
\beq\label{eq:exp}
N_B-1= m_j (N_j-1) + n_{ij} N_{i\neq j}.
\eeq
One notices that the inter-component correlations induce additional zeros in the wave function; this is energetically
favourable because the SU(4)-symmetric Coulomb interaction is as strong between particles of the same component as between
those belonging to different ones. Furthermore, one notices that Eq. (\ref{eq:exp}) has the character of a matrix equation, and it
turns out to be useful to introduce the exponent matrix $\Mmath=n_{ij}$ the diagonal elements of which are simply the
intra-component exponents $n_{jj}\equiv m_j$ and the off-diagonal ones those corresponding to inter-component correlations. In terms
of this exponent matrix, the relation between the component filling factors and the exponents reads
\beq\label{eq:CompFill}
\left(\begin{array}{c} \nu_1 \\ \nu_2 \\ \nu_3 \\ \nu_4 \end{array} \right) = \Mmath^{-1}
\left(\begin{array}{c} 1 \\ 1 \\ 1 \\ 1 \end{array} \right).
\eeq
In the zero-energy LL, the total filling factor is related to the component filling factors by
\beq\label{eq:nuDef}
\nu=-2+\sum_j^4 \nu_j,
\eeq
as a consequence of its half-filling for $\nu=0$, whereas in all other LLs $n$ the filling factor reads
\beq
\nu=\pm[4(n-1)+2] + \sum_j^4\nu_j.
\eeq
Notice that a state at a filling factor $\nu$ is related to another one at $-\nu$ by particle-hole symmetry. 

In addition to the determination of the component filling factors, the exponent matrix $\Mmath$ is useful also in two other
respects. First, it allows one to distiguish between potential physical states and those that cannot describe a homogeneous
liquid state that displays the FQHE. Indeed, the matrix must be positive definite, i.e. contain only positive (or zero) eigenvalues,
unless the corresponding state is unstable and undergoes a phase separation between the different components.\cite{DGRG}
Second, The matrix $\Mmath$ encodes prominent properties of the quasiparticle excitations, such as their fractional charge and
their statistics.\cite{WZ}
Finally, the rank of the matrix $\Mmath$ encodes the SU(4)-ferromagnetic properties of the different states.\cite{GR07}
In order to illustrate
this point, we first mention that Eq. (\ref{eq:CompFill}) is only well-defined if $\Mmath$ is invertible (of rank 4). This means that 
all component filling factors are fixed, and thus also all polarisations which are simple combinations of these factors; e.g.
the spin polarisation (in the $z$-direction) is simply given by $S_z=(N/2)(\nu_{\ua,K} + \nu_{\ua,K'} - \nu_{\da,K} - \nu_{\da,K'})$,
whereas the valley-pseudospin polarisation reads $P_z=(N/2)(\nu_{\ua,K} - \nu_{\ua,K'} + \nu_{\da,K} - \nu_{\da,K'})$.
As an example, one may invoke the state with $m_j=3$ for all $j$ and $n_{ij}=2$ for all $i\neq j$.\cite{toke07}
All component filling factors are fixed to be $\nu_j=1/9$, as may be seen from Eq. (\ref{eq:CompFill}), and this state would be 
an SU(4)-singlet candidate for a (yet unobserved) FQHE at $\nu=-2+4/9$.

In the opposite limit, where $\Mmath$ is of rank 1, the component filling factors are fully undetermined -- the only combination
that is fixed is the total sum. This is precisely the case of Laughlin's wave function, which may be described as 
a four-component Halperin wave function (\ref{eq:GenHalp}) with all $n_{ij}=m_j=m$ being the same odd integer. As we have already 
mentioned, this corresponds to a fully polarised SU(4)-ferromagnetic state, and the sum of component filling factors is
just $\sum_j^4\nu_j=1/m$.

There are intermediate states for which, e.g., only one of the polarisations is fixed. As an example, we may consider the
state with $m_j=3$, $n_{12}=n_{14}=n_{23}=n_{34}=3$, and $n_{13}=n_{24}=2$, which is described by an exponent matrix of rank 2. 
In addition to the total filling factor, which is fixed at $\nu=-2+2/5$, the combinations $\nu_1+\nu_3=1/5$ and $\nu_2+\nu_4=1/5$
are fixed. If we identify, for illustration reasons, the components as $\{1,2,3,4\}=\{(\ua,K),(\ua,K'),(\da,K),(\da,K')\}$,
this state would correspond to a valley-pseudospin singlet, with $\nu_K=\nu_{(\ua,K)}+\nu_{(\da,K)}=1/5$, 
$\nu_{K'}=\nu_{(\ua,K')}+\nu_{(\da,K')}=1/5$, such that $P_z=(N/2)(\nu_K-\nu_{K'})=0$, whereas the spin is polarised and
free to be oriented, e.g. by an external Zeeman effect. 

Interestingly, the 2/5 and 4/9 states discussed above are in competition with completely polarised CF states,
that may occur at filling factors $\nu=-2 + p/(2p+1)$, in terms of the integers $p$. Notice that the Halperin-type states at 
$2/5$ and $4/9$, which we discuss here, may alternatively be viewed as unpolarised CF states.\cite{toke07} 
Numerical calculations have shown that 
the polarised states are generally higher in energy in the zero-energy graphene LL,\cite{toke07} 
but they may become competitive
when external symmetry-breaking terms are taken into account, such that one may expect similar spin transistions 
at fixed filling factors as in 2D electron systems in GaAs heterostructures.\cite{2_5spinExp}
Quite generally, it is important to stress that the polarisation may change drastically when varying the filling factor --
even if the LL degeneracy may be lifted in a precise hierarchy at the integer filling factors $\nu=0$ and $\nu=\pm 1$, this
hierarchy is easily destroyed when shifting the filling factor away from these values, as a consequence of the dominant 
Coulomb interaction. From a theoretical point one therefore expects that a fully spin-valley polarised state at $\nu=-2+1/3$
is depolarised when increasing the filling factor -- this depolarisation is very efficient because of the low-energy skyrmion
excitations of the SU(4) ferromagnetism associated with the Laughlin state. Indeed, upon increase of $\nu$ 
one obtains a state at $\nu=-2+2/5$ that is
polarised in only one of the channels, e.g the spin for a Zeeman effect in the absence of valley-symmetry
breaking terms. Upon further increase, the spin-valley polarisation disappears completely at $\nu=-2+4/9$, where an SU(4) 
singlet is the state of lowest energy, in the absence of extrinsic symmetry-breaking effects. However, when approaching 
the filling factor $\nu=-1$, the SU(4) spin-valley polarisation is again expected to be fully restored. 

We emphasise that, apart from the above-mentioned states at $\nu=2/5$ and $4/9$, most of the Halperin states (\ref{eq:GenHalp})
are not eigenstates of the SU(4) symmetry group and thus not of the bare SU(4)-symmetric Coulomb interaction. However, as we
discuss in the following subsection for some of such states, they can be stabilised with the help of relatively weak
symmetry-breaking terms.
We finally notice that this picture and the expected polarisation of the discussed FQHE states also holds true, within
numerical calculations, fo the LL $n=1$.\cite{toke07}

\subsubsection{The 1/3 family of FQHE states}

\begin{figure}
\centering
\includegraphics[width=4.5cm,angle=0]{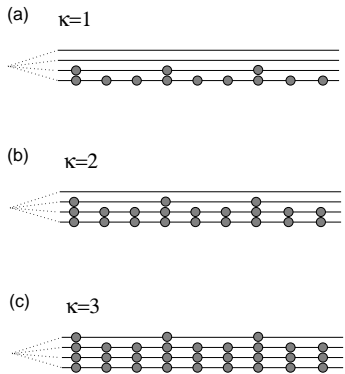}
\caption{\footnotesize{Sketch of different states $\psi_{\nu=-2+\kappa+1/3}$ of the 1/3 family, for 
\textit{(a)} $\kappa=1$, \textit{(b)} $\kappa=2$, and \textit{(c)} $\kappa=3$.
}}
\label{fig:02}
\end{figure}

We have discussed, until now, the use of Halperin wave functions for potential FQHE states
in the filling-factor range $-2<\nu<-1$. Notice, however, the the only $n=0$ FQHE states that have been clearly identified
ar $\nu=\pm 1/3,\pm 2/3$, and $\pm 4/3$, whereas the states at $\nu=\pm 5/3$ are absent or extremely 
weak.\cite{du09,bolotin09,ghahari10,dean10}
The understanding of these states requires the inclusion of fully occupied spin-valley subbranches in addition to partially
filled ones. In principle, these states may also be described in terms of generalised Halperin wave functions with 
broken SU(4) symmetry, and we terminate this review with a brief discussion of them. The states, which are depicted in Fig. \ref{fig:02},
may be constructed by the wave functions 
\beq\label{eq:1_3WF}
\psi_{\nu=-2+\kappa+1/3}=\prod_{j}^{\kappa}\prod_{k_j<l_j}^{N_B}\left(z_{k_j}^{(j)}-z_{l_j}^{(j)}\right)\prod_{k<l}^{N_B/3}
\left(w_{k}-w_{l}\right)^3,
\eeq
where $z_{k_j}^{(j)}$ are the complex coordinates of $j$-type particles that reside in the $\kappa$ fully occupied spin-valley
branches $j=1,...,\kappa$ (since $\nu_j=1$, we have $N_j=N_B$), whereas $w_k$ is that of a particle in the other spin-valley branches, 
occupied by $N_B/3$ particles, where we do not specify explicitly the component. This state is described by an exponent matrix
with $m_j=1$ for $j=1,...,\kappa$, $m_j=3$ for $j=\kappa+1, ...,4$, $n_{ij}=0$ if one of the indices is $1,...,\kappa$ and 
$n_{ij}=3$ otherwise. In the framework of quantum Hall ferromagnetism, the state may be viewed as $\kappa$ ``inert'' 
levels,\footnote{They are not really inert because they are responsible for low-energy spin-flip excitations, as we discuss
below.} 
whereas the electrons in the $4-\kappa$ partially occupied subbranches form a Laughlin-type state with incorporated SU($4-\kappa$)
ferromagnetic low-energy excitations in terms of $(3-\kappa)$-fold degenerated spin-valley waves.
Other members of the 1/3 family may be obtained from the states (\ref{eq:1_3WF}) with the help of a 
particle-hole transformation.

As discussed in the previous section, the states (\ref{eq:1_3WF}) cannot describe the ground state of the Coulomb interaction
because they are no eigenstates of the SU(4) symmetry. These states may be stabilised artificially by a particular choice of the interaction potential between the different particles that explicitly breaks the SU(4) symmetry.\cite{papic10} 
However, the more physical approach which we adopt here shows that the states (\ref{eq:1_3WF}) may be relevant 
even for an SU(4)-symmetric interaction potential as soon as the extrinsic symmetry-breaking terms are included.

\begin{figure}
\centering
\includegraphics[width=8.5cm,angle=0]{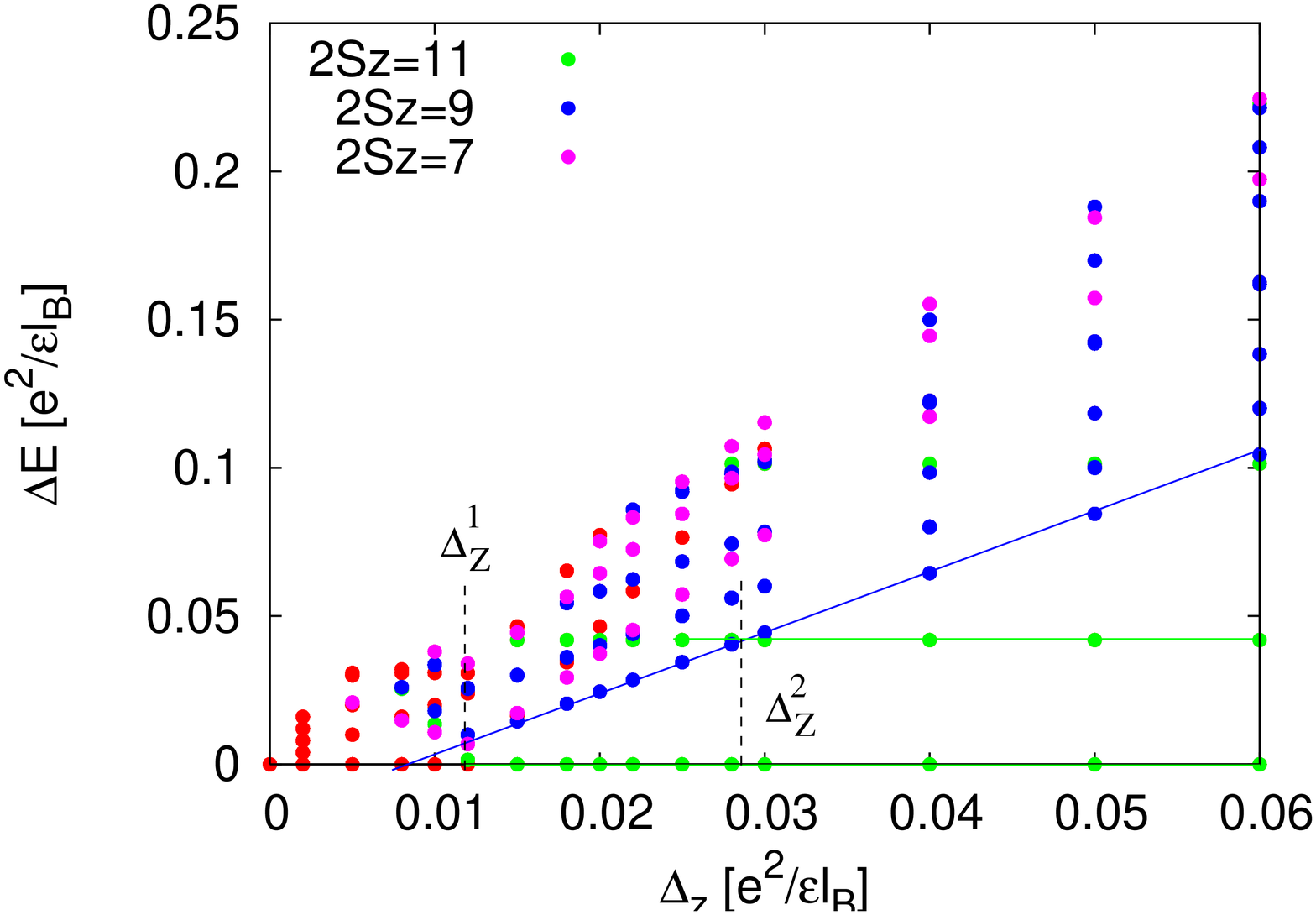}
\caption{\footnotesize{Energy spectrum obtained in Ref. \onlinecite{papic10} by exact diagonalisation of
$N=17$ electrons interacting via the Coulomb interaction on a sphere threaded by $N_B=6$ flux quanta. The spectrum
is shown as a function of the Zeeman effect, in units of $e^2/\epsilon l_B$. The colours indicate levels with
the different spin polarisations $S_z=11/2$ (green, corresponding to the state $\psi_{1/3}$) $9/2$ (blue) adn $7/2$ (pink).
Red dot correspond to energy levels with other polarisation.  
}}
\label{fig:03}
\end{figure}

For $\kappa=2$ [see Fig. \ref{fig:02}(b)],
i.e. for a filling factor $\nu=1/3$, the energy spectrum obtained from exact diagonalisation is
shown in Fig. \ref{fig:03} as a function of an extrinsic symmetry-breaking field that has been chosen to be the Zeeman
effect. The spectrum has been obtained with the help of the DiagHam\cite{DiagHam} code with implemented SU(4) 
symmetry for up to $N=17$ electrons on a sphere threaded by $N_B= 6$ flux quanta.
As expected from the above discussion, the ground state in the absence of a Zeeman effect is not the state
$\psi_{\nu=1/3}$ described by Eq. (\ref{eq:1_3WF}) because it does not have the correct spin polarisation.
However, the state is stabilised already for very small symmetry breaking, i.e. for a Zeeman effect above
$\Delta_Z^1\simeq 0.01 e^2/\epsilon l_B$, as may be seen in Fig. \ref{fig:03}, where the ground state has the correct 
(maximal) spin polarisation $S_z=11/2$ (green dots). Although the state $\psi_{\nu=1/3}$
is the ground state above this critical value of the Zeeman effect,
its low-energy excitations are not the usual collective excitations of the Laughlin state, but coherent spin-flip excitations
in the sector $S_z=9/2$ that are represented by the blue line in Fig. \ref{fig:03}. These spin-flip excatiations are the
relevant modes below a second critical Zeeman field $\Delta_Z^2\simeq 0.03 e^2/\epsilon l_B$, whereas
above $\Delta_Z^2$ the lowest-energy
excitations are the usual charge excitations in the same polarisation sector.

The case $\kappa=1$, which corresponds to a single fully occupied spin-valley branch [see Fig. \ref{fig:02}(a)],
may be checked within a simplified 
two-component scheme that neglects the two-fold degenerate spin-valley modes associated with the partially filled three subbranches 
$j=2$, 3 and 4. The corresponding (simplified) wave function reads
\beq\label{eq:simpl1_3}
\psi_{\nu=-1+1/3}^{\rm 2-comp}=\prod_{k<l}^{N_B}\left(z_{k}-z_{l}\right)\prod_{k<l}^{N_B/3}
\left(w_{k}-w_{l}\right)^3,
\eeq
where $z_k$ is the position of a particle in the fully occupied subbranch and $w_k$ one in the 1/3-filled second component. This
wave function has been tested in exact-diagonalisation calculations with the help of the DiagHam code\cite{DiagHam} with
an implemented SU(2)-symmetric Coulomb potential, for $N=22$ electrons on a sphere with $N_B=15$ flux quanta.\cite{papic10} 
The obtained energy spectrum shows the same features as that depicted in Fig. \ref{fig:03} obtained within a four-component
calculation, with a similar critical field $\Delta_Z^1\simeq 0.01 e^2/\epsilon l_B$ above which the state (\ref{eq:simpl1_3})
is stabilised albeit with a slightly larger field $\Delta_Z^2\simeq 0.08 e^2/\epsilon l_B$, below which the collective 
excitations are dominated by spin-flip excitations.\cite{papic10} 

These two results obtained numerically, for $\kappa=2$ and for $\kappa=1$ in a simplified version, hint at a certain
universality in the mechanism of stabilising states of the form (\ref{eq:1_3WF}) by weak extrinsic symmetry-breaking fields. 
The physical picture that emerges from it may be summarised  as follows: the 
systems has an interaction-driven tendency to form such states (here the states $\psi_{\nu=-2+\kappa+1/3}$ of the 1/3 family), but
a small extrinsic SU(4) spin-valley symmetry breaking is nevertheless necessary to stabilise them. This needs to be contrasted 
to the SU(4) quantum Hall ferromagnetism discussed in Sec. \ref{sec:QHFM} where the state remains stable even in the complete absence
of extrinsic symmetry-breaking effects. 

We finally notice that, even in the intermediate regime $\Delta_Z^1<\Delta_Z<\Delta_Z^2$, 
the lowest-energy excitations in the limit $ql_B\gg 1$ are not the collective spin-flip excitations, but as for the usual Laughlin 1/3 
state quasi-particle excitations that may eventually be dressed by SU($4-\kappa$) spin-valley textures in the partially occupied
subbranches and that are responsible for the activation gap measured in the experiments.\cite{ghahari10,dean10}

\section{Conclusions}

In conclusion, we have reviewed theoretically the role of electronic interactions in graphene Landau levels. These interactions
are responsible for two prominent effects: (a) the formation of SU(4) spin-valley quantum Hall ferromagnets that are likely
to be responsible for the observed IQHE at $\nu=0$, $\pm 1$, $\pm 3$, $\pm 4$, and $\pm 5$ that do not belong to the series
(\ref{eq:RQHE}) of the usual (relativistic) graphene IQHE; and (b) the recently observed FQHE. Although both effects are known
also in the context of non-relativistic quantum Hall systems, such as in GaAs heterostructures, they are different in graphene
as a consequence of the approximate SU(4) spin-valley symmetry of the Coulomb interaction potential. Even if the SU(4)
symmetry of graphene LLs is broken by extrinsic effects, such as the Zeeman effect or a valley-pseudospin Zeeman-type effect 
due to static lattice distortions in graphene, the latter effects are associated with energy scales that are much smaller 
than the leading Coulomb interaction scale, for physically accessible magnetic fields. These extrinsic effects are mainly
cooperative with the tendency of forming maximally spin-valley polarised states. In the context of quantum Hall ferromagnetism,
they orient the preformed spin-valley magnetisation into particular channels, whereas they are necessary to stabilise 
the trial states $\psi_{\nu=-2+\kappa+1/3}$ that may account for the experimentally observed members of the 1/3 family.

Also other FQHE states than those of the above-mentioned 1/3 family may be described in the framework of the SU(4) theory
of the FQHE and are expected to display very special spin-valley polarisations. It remains an experimental challenge to have 
access to these states and their physical properties, but from an experimental point of view we seem to be only at the beginning
of the discovery of the possibly very rich physical properties 
of the graphene FQHE. The expected findings of novel FQHE states in graphene
may provide other surprises that will certainly also challenge the SU(4) theory of the graphene FQHE.

\section*{Acknowledgments}

We acknowledge the collaboration on multi-component FQHE with Zlatko Papi\'c and Rapha\"el de Gail.
Deep insight has been obtained within their PhD and Master studies, respectively. 
Furthermore, we acknowledge Beno\^it Dou\c cot, Roderich Moessner, and Pascal Lederer for their
collaboration on the understanding of SU(4)-quantum-Hall ferromagnetism as well as Jean-No\"el Fuchs and Rafael Rold\'an
for that on electronic interactions and collective excitations in the IQHE regime. Finally, very stimulating discussions
with Philip Kim need to be acknowledged that provided experimental guidance to understanding the SU(4) FQHE. This work was funded by
Agence Nationale de la Recherche under Grant Nos. ANR-JCJC-0003-01, ANR-06-NANO-019-03, and ANR-09-NANO-016.

\end{document}